\documentclass[11pt]{article}

\usepackage[draft]{graphics}
\usepackage{latexsym}

\newcommand{\be}{\begin{equation}}
\newcommand{\ee}{\end{equation}}

\newcommand{\ba}{\begin{eqnarray}}
\newcommand{\ea}{\end{eqnarray}}

\newcommand{\bat}{\begin{tabular}{lr}}
\newcommand{\eat}{\end{tabular}}

\newcommand{\bay}{\[\begin{array}{lr}}
\newcommand{\eay}{\end{array}\]}

\newcommand{\baa}{\begin{eqnarray*}}
\newcommand{\eaa}{\end{eqnarray*}}

\def\br{{\mathbf{r}}}

\def\sg{{\sigma}}
\def\ros{{\rho(\sg)}}

\def\ag{{\alpha}}

\begin{document}
\bibliographystyle{prbsty}
 
\title{Density functional theory of phase coexistence in\\
weakly polydisperse fluids} 
 
\author{Hong Xu $^{\dag}$ and Marc Baus $^{\dag\dag}$}
\date{}
\maketitle

\noindent
$^{\dag}$ D\'epartement de Physique des Mat\'eriaux (UMR 5586 du CNRS),\\
Universit\'e Claude Bernard-Lyon1, 69622 Villeurbanne Cedex, France\\
$^{\dag\dag}$ Physique des Polym\`eres, Universit\'e Libre de Bruxelles,\\
Campus Plaine, CP 223, B-1050 Brussels, Belgium\\

\vspace{2truecm}
\pagebreak
 
\begin{abstract}
The recently proposed universal relations between the moments
of the polydispersity distributions of a phase-separated weakly
polydisperse system are analyzed in detail using the numerical
results obtained by solving a simple density functional theory
of a polydisperse fluid. It is shown that universal properties
are the exception rather than the rule.\par
\end{abstract} 

\pagebreak

\vspace{1truecm}
Many, natural or man-made, systems are mixtures of similar instead of
identical objects. For example, in a colloidal dispersion\cite{rus}
the size and surface charge of the colloidal particles are usually
distributed in an almost continuous fashion around some mean value. When
this distribution is very narrow the system can often be assimilated
\cite{poon} to a one-component system of identical objects.
Such a system is usually called monodisperse whereas otherwise
it is termed polydisperse. Since polydispersity is a direct consequence
of the physico-chemical production process it is an intrinsic
property of many industrial systems. Therefore, many author\cite{laro}
have included polydispersity into the description of a given phase
of such systems. More recently, a renewed interest can be witnessed
for the study of phase transitions occuring in weakly polydisperse
systems\cite{evan}. The phase behavior of polydisperse systems is
of course much richer than that of its monodisperse counterpart. It is
also more difficult to study theoretically, essentially because
one has to cope with an infinity of thermodynamic coexistence
conditions\cite{laro}. Therefore, several authors have proposed
approximation schemes\cite{sol} which try to bypass this difficulty.
In the present study we take to opposite point of view by solving
numerically the infinitely many thermodynamic coexistence conditions
for a simple model polydisperse system. On this basis we have studied
the radius of convergence of the weak polydispersity expansion
used in ref.4 and found that their ``universal law of fractionation" and
some of their conclusions have to be modified in several cases.\par

The statistical mechanical description of a polydisperse equilibrium
system is equivalent to a density functional theory\cite{han} for a
system whose number density, $\rho(\br,\sg)$, depends besides the
position variable $\br$ (assuming spherical particles) also on
at least one polydispersity variable $\sg$ (which we consider to be
dimensionless). Such a theory is completely determined once the
intrinsic Helmholtz free-energy per unit volume, $f[\rho]$, has been
specified as a functional of $\rho(\br,\sg)$ (for notational
convenience the dependence on the temperature $T$ will not be
indicated explicitly). For the spatially uniform fluid phases
considered here (and also implicitly in ref.4) we have,
$\rho(\br,\sg)\rightarrow\ros$, and the pressure can be written
as, $p[\rho]=\int d\sg\ros\mu(\sg;[\rho])-f[\rho]$, where
$\mu(\sg;[\rho])=\delta f[\rho]/\delta \ros$, is the chemical potential of
``species" $\sg$. When a parent phase of density $\rho_0(\sg)$
phase separates into $n$ daughter phases of density $\rho_i(\sg)$
($i=1,\ldots,n$) the phase coexistence conditions imply that,
$p[\rho_1]=p[\rho_2]=\ldots=p[\rho_n]$, and 
$\mu(\sg;[\rho_1])=\mu(\sg;[\rho_2])=\ldots=\mu(\sg;[\rho_n])$.
For simplicity we consider here only the case of two daughter
phases ($n=2$) and rewrite moreover $\rho_i(\sg)=\rho_i h_i(\sg) (i=0,1,2)$
in terms of the average density $\rho_i$ and a polydispersity
distribution $h_i(\sg)$ such that $\int d\sg h_i(\sg)=1$.
Since the ideal gas contribution to $f[\rho]$ is exactly known\cite{han} 
one has, $\mu(\sg;[\rho])=k_B T\ln\{\Lambda^3(\sg)\ros\}
+\mu_{ex}(\sg;[\rho])$, where $k_B$ is Boltzmann's constant,
$\Lambda(\sg)$ is the thermal de Broglie wavelength of species
$\sg$ and $\mu_{ex}$ the excess (ex) contribution to $\mu$.
This allows us to rewrite the equality of the chemical potentials
of the two daughter phases as, $h_1(\sg)=h_2(\sg)A(\sg)$, where
$A(\sg)$ is a shorthand notation for:
\be
A(\sg)=\frac{\rho_2}{\rho_1}\exp\beta\left\{\mu_{ex}(\sg;[\rho_2])
-\mu_{ex}(\sg;[\rho_1])\right\}
\ee
with $\beta=1/k_B T$. The polydispersity distributions are further
constrained by the relation, $x_1 h_1(\sg)+x_2 h_2(\sg)=h_0(\sg)$,
which expresses particle number conservation. The number
concentration of phase 1, $x_1=1-x_2$, is
given by the lever rule: 
$x_1=\frac{\rho_1}{\rho_1-\rho_2}\cdot\frac{\rho_0-\rho_2}{\rho_0}$.
Combining these two relations one finds:
\be
h_2(\sg)-h_1(\sg)=h_0(\sg)\cdot H(\sg)
\ee
where $H(\sg)\equiv (1-A(\sg))/(x_2+x_1 A(\sg))$. Eq.(2) is the starting
point to relate the difference between the moments 
of the daughter phases, $\Delta_k=\int d\sg\,\sg^k(h_2(\sg)-h_1(\sg))$,
to the moments, $\xi_k=\int d\sg\,\sg^k h_0(\sg)$ ($k=0,1,2,\ldots$), of
the parent phase distribution $h_0(\sg)$. Indeed, when $\sg$ is
chosen such that $h_0(\sg)$ tends to the Dirac delta function $\delta(\sg)$
in the monodisperse limit, $\Delta_k$ can be obtained from (2) by
expanding $H(\sg)$ around $\sg=0$, $H(\sg)=\sum_{l=0}^{\infty} a_l\sg^l$,
yielding for a weakly polydisperse system, 
$\Delta_k=\sum_{l=0}^{\infty} a_l\xi_{l+k}$. 
The normalization of the $h_i(\sg)$ ($i=0,1,2$) implies
$\Delta_0=0$, $\xi_0=1$ or $a_0=-\sum_{l=1}^{\infty}a_l\xi_l$,
and eliminating $a_0$ from $\Delta_k$  
yields the general moment relation:
\be
\Delta_k=a_1\xi_{k+1}+\sum_{l=2}^{\infty} a_l(\xi_{k+l}-\xi_l\xi_k).
\ee
where we took moreover into account that
$\sg$ can always be chosen such that $\xi_1=0$.
When only the first term in the r.h.s. of (3) is retained we recover
the universal law $\Delta_k/\Delta_l=\xi_{k+1}/\xi_{l+1}$, put
forward in ref.4. The question left unanswered by the study of
ref.4 concerns the radius of convergence of the weak polydispersity
expansion (3). In order to study this problem in more detail
we now consider
a simple model system for which we can determine the $h_i(\sg)(i=1,2)$
numerically and compare the results with (3). The free energy
density functional chosen here corresponds to a simple van der Waals
(vdW) model\cite{gua} for the liquid-vapor transition in polydisperse
systems of spherical particles of variable size:
\ba
f[\rho]=&k_B\,T\int d\sg\ros\left\{\ln(
\frac{\Lambda^3(\sg)\ros}{E[\rho]})-1\right\}\nonumber\\
&+\frac{1}{2}\int d\sg\int d\sg'\,V(\sg,\sg')\ros\rho(\sg')
\ea
where, $E[\rho]=1-\int d\sg\,v(\sg)\ros$, describes the average excluded volume correction for
particles of radius $R_\sg$ and volume $v(\sg)=\frac{4\pi}{3}R_\sg^3$,
while $V(\sg,\sg')=\int d\br\,V(r;\sg,\sg')$ is the integrated attraction
between two particles of species $\sg$ and $\sg'$, for which we took
the usual vdW form, $V(r;\sg,\sg')=-\epsilon_0(R_\sg+R_{\sg'})^6/r^6$
for $r\ge R_\sg+R_{\sg'}$ and zero otherwise, $\epsilon_0$ being the
amplitude of the attraction at the contact of the two particles.
The size-polydispersity can be described
in terms of the dimensionless variable, $\sg=R_\sg/R\,-1$, with
$R$ the mean value of $R_\sg$ in the parent phase, hence
$\xi_1=\int d\sg\,\sg h_0(\sg)=0$. The thermodynamics is given in terms
of $h_0(\sg)$, the dimensionless temperature $t=k_B\,T/\epsilon_0$ and the dimensionless density $\eta=v_0\rho$, with
$v_0=\frac{4\pi}{3}R_0^3$ and $R_0$ the value of $R_\sg$ 
in the monodisperse limit.
The coexistence conditions are integral equations which can be solved
numerically using, for instance, an iterative algorithm\cite{mix}
for any $t$, $\eta_0=v_0\rho_0$ and $h_0(\sg)$. For $h_0(\sg)$ we took
a Schulz distribution\cite{laro} with zero mean. The normalized 
distribution is given, for $-1\le\sg<\infty$, by 
$h_0(\sg)=\ag^\ag(1+\sg)^{\ag-1}e^{-\ag(1+\sg)}/\Gamma(\ag)$,
with $\Gamma(\ag)$ the gamma function and $1/\ag$ a width parameter which
measures the distance to the monodisperse limit, $h_0(\sg)\rightarrow\delta(\sg)$
when $\ag\rightarrow\infty$. We then have: $\xi_0=1$, $\xi_1=0$, $\xi_2=1/\ag$,
$\xi_3=2/\ag^2$, $\xi_4=\frac{3}{\ag^2}+\frac{6}{\ag^3}$,
$\xi_5=\frac{20}{\ag^3}+\frac{24}{\ag^4}$, etc. For a weakly polydisperse
system we retain only the dominant terms of (3) in a $1/\ag$ expansion.
From (3) we obtain then:
$\Delta_1=a_1(\infty)\xi_2+O(1/\ag^2)$,
$\Delta_2=a_1(\infty)\xi_3+a_2(\infty)(\xi_4-\xi_2^2)+O(1/\ag^3)
=\{a_1(\infty)+a_2(\infty)\}\xi_3+O(1/\ag^3)$,
$\Delta_3=a_1(\infty)\xi_4+O(1/\ag^3)$, etc, where $a_l(\infty)$ are
the values of $a_l$ for $\ag\rightarrow\infty$.   
Using the vdW expression (4) to evaluate (1) one finds, for ex. for
$t=1.0$ and $\eta_0=0.5$, $a_1(\infty)=1.75$ and $a_2(\infty)=-2.68$.
Using the corresponding numerical solutions found for $h_1(\sg)$ and
$h_2(\sg)$ (see Fig.1) it can be seen from Fig.2 that 
$\Delta_1/\xi_2\approx 1.75$, $\Delta_2/\xi_3\approx -0.93$ and $\Delta_3/\xi_4\approx 1.75$ are
obeyed to within ten percent for $\ag$ larger than, respectively,
40, 80 and 150.
We can conclude thus that the weak
polydispersity expansion (3) is valid (to dominant order) for Schulz
distributions $h_0(\sg)$ with a dispersion $\left((\xi_2-\xi_1^2)^{1/2}\right)$
smaller than, say, 0.1 ($\ag\approx 100$). These values do of course depend on the thermodynamic
state but the case considered here ($t=1$, $\eta_0=0.5$) is representative
of other $t,\eta_0$ values. Note also that we have verified 
numerically that the radius of convergence of (3) with respect to
$1/\ag$ is fairly sensitive to the total amount of polydispersity
present. Allowing, for instance, the amplitude $\epsilon_0$ of the pair
potential $V(r;\sg,\sg')$ to depend on $\sg$ and $\sg'$ does reduce the
radius of convergence of (3) considerably. From the above it follows that,
$\frac{\Delta_3}{\Delta_1}$ follows the universal law, 
$\frac{\Delta_3}{\Delta_1}=\frac{\xi_4}{\xi_2}$, put forward in
ref.4 whereas $\frac{\Delta_2}{\Delta_1}$ follows the non-universal
law, $\frac{\Delta_2}{\Delta_1}
=\{1+\frac{a_2(\infty)}{a_1(\infty)}\}\frac{\xi_3}{\xi_2}$.
We have verified that similar results can be obtained for different
$h_0(\sg)$ distributions. Taking, for instance, a Gaussian for
$h_0(\sg)$ similar results are found although $\xi_3=0$
for this case. This invalidates the conclusion of ref.4 that a particular
importance should be attached to the skewness of $h_0(\sg)$. In conclusion,
the general moment relation (3) can yield useful information about the
phase behavior of weakly polydisperse systems but this information
is in general not universal.\par
\pagebreak
\pagebreak
\noindent{\bf Figure Captions}\par
\vspace{1truecm}
\noindent
{\bf FIG. 1.} The polydispersity distributions $h_n(\sg)$ of the
parent phase ($n=0$: full curve) (a Schulz distribution with the width parameter
$\ag=50$), the low-density (n=1: dotted curve) and the high-density (n=2: circles) daughter
phases, 
as obtained by numerically solving the coexistence conditions
of the van der Waals model of eq.(4) for $t=1$, $\eta_0=0.5$.
The corresponding dimensionless densities of the coexisting
daughter phases are $\eta_1=0.106$, $\eta_2=0.521$ 
whereas for the monodisperse
system one has, $\eta_1=0.103$, $\eta_2=0.608$. Also shown are $h_1(\sg)-h_0(\sg)$ 
(dashed curve) and $[h_2(\sg)-h_0(\sg)]\cdot 50$ (triangles).\par
\vspace{0.3truecm}
\noindent
{\bf FIG. 2.} The ratio $\Delta_k/\xi_{k+1}$ ($k=1,2,3$) versus $1/\ag$
as obtained from the numerical solution of the van der Waals model
of eq.(4) for $t=1$, $\eta_0=0.5$ and a Schulz distribution
for $h_0(\sg)$. The symbols are as follows:  circles(k=1),
squares(k=2) and triangles(k=3). 
The dotted lines indicate their asymptotic
($\ag\rightarrow\infty$) values. The arrows indicate
for each case the
radius of convergence of the weak polydispersity expansion of eq.(3). \par

\end{document}